# Nontrivial coupling of light into a defect: the interplay of nonlinearity and topology

Shiqi Xia[1+], Dario Jukić[2+], Nan Wang[1+], Daria Smirnova[3], Lev Smirnov[4], Liqin Tang[1,5], Daohong Song[1,5], Alexander Szameit[6], Daniel Leykam[7,8], Jingjun Xu[1,5], Zhigang Chen[1,5,9], and Hrvoje Buljan[1,10]

[1] *The MOE Key Laboratory of Weak-Light Nonlinear Photonics, TEDA Applied Physics Institute and School of Physics, Nankai University, Tianjin 300457, China*
[2] *Faculty of Civil Engineering, University of Zagreb, A. Kačića Miošića 26, 10000 Zagreb, Croatia*
[3] *Nonlinear Physics Centre, Research School of Physics, Australian National University, Canberra ACT 2601, Australia*
[4] *Institute of Applied Physics, Russian Academy of Science, Nizhny Novgorod 603950, Russia*
[5] *Collaborative Innovation Center of Extreme Optics, Shanxi University, Taiyuan, Shanxi 030006, People's Republic of China*
[6] *Institut für Physik, Universität Rostock, Albert-Einstein-Strasse 23, 18059 Rostock, Germany*
[7] *Center for Theoretical Physics of Complex Systems, Institute for Basic Science (IBS), Daejeon 34126, Korea*
[8] *Basic Science Program, Korea University of Science and Technology, Daejeon 34113, Korea*
[9] *Department of Physics and Astronomy, San Francisco State University, San Francisco, California 94132, USA*
[10] *Department of Physics, Faculty of Science, University of Zagreb, Bijenička c. 32, 10000 Zagreb, Croatia*

[+]These authors made equal contribution.
*songdaohong@nankai.edu.cn*, *hbuljan@phy.hr*, *zgchen@nankai.edu.cn*

**Abstract:**

**The flourishing of topological photonics in the last decade was achieved mainly due to developments in linear topological photonic structures. However, when nonlinearity is introduced, many intriguing questions arise. For example, are there universal fingerprints of underlying topology when modes are coupled by nonlinearity, and what can happen to topological invariants during nonlinear propagation? To explore these questions, here we experimentally demonstrate nonlinearity-induced coupling to topologically protected edge states using a photonic platform, and theoretically develop a general framework for interpreting the mode-coupling dynamics in nonlinear topological systems. Performed in laser-written photonic Su-Schrieffer-Heeger lattices, our experiments reveal nonlinear coupling of light into a nontrivial edge or interface defect channel otherwise not permissible due to topological protection. Our theory explains well all the observations. Furthermore, we introduce the concepts of inherited and emergent nonlinear topological phenomena, and a protocol capable of unveiling the interplay of nonlinearity and topology. These concepts are applicable for other nonlinear topological systems, either in higher dimensions or beyond our photonic platform.**

## Introduction

Topological photonics has become one of the most active research frontiers in optics over the last decade[1, 2]. The initial ideas were drawn from condensed matter physics, where the concept of topology was found crucial for understanding of the celebrated quantum Hall effect (QHE)[3, 4] and, later on, for the development of topological insulators[5-7]. In 2008, Raghu and Haldane proposed that the Bloch bands of photonic crystals designed with time-reversal symmetry-breaking elements can have nontrivial topological invariants[8, 9], namely, the non-zero Chern numbers. When two materials with different topological invariants are interfaced, bulk-edge correspondence[2, 10, 11] guarantees the existence of topological edge states, which enjoy robust unidirectional propagation. Such correspondence holds in both quantum and classical wave systems, which inspired the first observation of unidirectional propagation of electromagnetic waves in the microwave regime[12]. Topological states of light and related phenomena were later realized in various systems, including photonic lattices[13], ring resonators[14] and metamaterials[15] (see Ref. [2] for a recent review[2]).

In electronic systems, the interplay of topology and quantum many-body interactions can result in intriguing topological states of matter such as the fractional QHE[4, 16]. An analogous, yet distinct avenue of research is to address the interplay of topology and nonlinearity in photonics. In conventional linear systems, the amount of energy present in each eigenmode remains constant during time evolution. When nonlinearity is introduced, however, it shuffles the energy between the eigenmodes, which brings memory back to the pioneering numerical experiment by Fermi, Pasta, Ulam and Tsingou who studied the thermalization induced by nonlinear coupling in 1955[17]. Their discovery of recurrence to a state very close to the initial condition in a surprisingly short time is rooted in the underlying integrability of the system. Such recurrence was recently observed with nonlinear optical spatial waves[18]. It is natural to wonder whether the eigenmodes of a topological system can be coupled by nonlinearity and, if so, how the nontrivial topology can be reflected onto subsequent dynamics, especially on the coupling of topologically protected edge states.

Thus far, nonlinear topological effects have been investigated far less than their linear counterparts, although nonlinearity inherently exists in many topological photonic platforms such as waveguide arrays, coupled resonators, and metamaterials [19-31]. Motivated by the unique functionalities and device applications, research in nonlinear topological photonics has been focused mainly on edge solitons in topological structures[21, 23, 32-34], nonlinearity-induced topological transitions[24, 25], nonlinear frequency generation[35-37], and topological lasing[38-40]. Despite these efforts, the fundamental issue of nonlinear coupling of eigenmodes in topological systems remains largely unexplored.

Here, we demonstrate nonlinearity-induced coupling to topologically protected edge states using a photonic platform and develop a general theoretical framework for interpreting the mode-coupling dynamics in nonlinear topological systems. Experimental results are obtained in photonic Su-Schrieffer-Heeger (SSH) lattices[41] fabricated with laser-writing technique in a nonlinear crystal. We observe that only under nonlinear excitation can a light beam travelling from the bulk to the edge of a nontrivial SSH lattice be coupled to the topologically protected edge state. Furthermore, nonlinear interaction of two beams at opposite incident angles is also observed, coupling into a topological interface state depending strongly on their relative phase. Our theory explains well these observations: under proper nonlinear excitation, the profile of the beam propagating along the edge (or interface) waveguide is inherited from that of the underlying linear topological system – overlapping above 98% with the linear topological edge states and with propagation constants residing in the band gap. When the nonlinearity is stronger than a critical value, however, the nonlinear eigenvalue of the edge state moves out of the gap and emerges above the first band, indicating that the localization is now dominated by the nonlinearity. The concepts introduced in this paper are generally applicable for nonlinear topological systems.

The SSH lattice exhibits two topologically distinct (Zak) phases, representing a prototypical one-dimensional (1D) topological system with chiral symmetry[2, 41, 42]. The SSH models have been implemented in a variety of platforms, including photonics and nanophotonics[43-49], plasmonics[50, 51], quantum optics[52-55], and particularly in the context of topological lasing[38, 56-58]. Such SSH-type models with driven nonlinearity have also attracted great attention[19, 24, 30, 32, 34-36, 59]. In particular, nonlinearity has been employed for spectral tuning[30] and time-domain pumping[59] of topological edge states, and for generation of topological gap solitons[32, 34] in such systems.

## Results

We study propagation of light in photonic lattices with a refractive-index variation given by $n_0 + \delta n_L(\mathbf{x}) + \delta n_{NL}(|\psi|^2)$, where $n_0$ is the constant part of the material's index of refraction, $\delta n_L(\mathbf{x})$ describes the linear photonic lattice which is uniform along the propagation axis $z$, and $\delta n_{NL}(|\psi|^2)$ is the nonlinear index change which depends on the intensity of the light (with $\psi(\mathbf{x}, z)$ being the complex amplitude of electric field). In the paraxial approximation, the propagation of light is modelled by the following Schrödinger-type equation with a nonlinear term:

$$i\frac{\partial\psi}{\partial z} = -\frac{1}{2k_0}\nabla^2\psi - \frac{k_0\delta n_L(\mathbf{x})}{n_0}\psi - \frac{k_0\delta n_{NL}(|\psi|^2)}{n_0}\psi(\mathbf{x},z) = (K + V_L + V_{NL})\psi, \qquad (1)$$

where we identify the kinetic term $K$, the linear index potential $V_L$ from $\delta n_L(\mathbf{x})$, and the nonlinear index potential $V_{NL}$ due to $\delta n_{NL}(|\psi|^2)$; $k_0$ is the wavenumber of light in the medium. The above equation holds for both 1D and 2D photonic lattices. In 1D systems, the spatial coordinate is a scalar $x$, and in 2D systems it is a vector $\mathbf{x} = x\hat{x} + y\hat{y}$. Here we consider a 1D linear topological system, that is, we assume that the photonic lattice $V_L$ can have nontrivial topological invariants. In our experiments and numerical simulations, we shall use the SSH lattice for $V_L(x)$. The photonic lattice and excitation scheme are illustrated in Figure 1, where Fig. 1(a1) corresponds to a nontrivial lattice (Zak phase $\pi$) with two topological edge modes in the gap, and Fig. 1(c1) corresponds to a trivial lattice (Zak phase 0) without any edge state. In our theory we use the above continuum model to describe the wave dynamics rather than its discrete version for better correspondence with experiments.

In our experiment, the 1D SSH photonic lattice as illustrated in Fig. 1 is established by the *continuous-wave* (CW) laser-writing technique, which writes the waveguide lattice site-to-site in the bulk of a 20-mm-long nonlinear photorefractive crystal[60]. Such a technique allows for inducing a topological defect not only at the edge [Fig. 1(a1)], but also at the center forming an interface [Fig. 2(a1)]. Different from the femtosecond-laser writing in fused silica[61], the lattice written in the nonlinear crystal is reconfigurable so it can be readily changed from trivial to nontrivial structures in the same crystal. Once a chosen structure is written, it remains invariant during the period of experimental measurements [see Methods]. In fact, since the SSH lattice is established here in the nonlinear crystal, it provides a convenient platform to investigate nonlinear wave dynamics in such a topological system, where the photorefractive nonlinear index potential $V_{NL}$ is easily controlled by a bias field and the beam intensity[19, 62]. Below we demonstrate nonlinearity-induced coupling to topologically protected states in two different cases.

In the first case, the topological defect is located at the SSH lattice edge (Fig. 1, left panels). When a narrow stripe beam (FWHM 12μm; input power 2.5μW) is launched straight into the edge waveguide under linear condition (the beam itself does not exhibit nonlinear self-action when the bias field is turned off), it evolves into a topological edge state (Fig. 1a2). Such an edge state, with characteristic amplitude and phase populating only the odd-number waveguides counting from the edge, is topologically protected by the chiral symmetry of the SSH lattice[2], as previously observed in the 1D photonic superlattice[43]. On the other hand, when the excitation is shifted away from the edge with a tilted broad beam to pump the defect ($k_x = 1.4\pi/a$, where $a$=38μm is the lattice constant), we observe that the beam does not couple into the edge channel under linear condition (Fig. 1b1). However, when the beam experiences a self-focusing nonlinearity (at a bias field of 160kV/m), a significant portion of the beam is coupled into the edge channel (Fig. 1b2), indicating that the nonlinearity somehow enables the energy flow from the bulk modes into the topological edge mode of the SSH lattice. According to Eq. (1), we perform numerical simulation to examine the nonlinear beam dynamics using parameters from the experiments, and the results are shown in Fig. 1(b3). Clearly, we see nonlinear coupling of the beam into the topological edge state of the SSH lattice, in agreement with experiment.

For direct comparison, in the right panels of Fig. 1, we present corresponding results obtained with the trivial SSH lattice. Dramatic difference is observed: (1) Under straight excitation, the input beam transports to quite a few waveguides close to the edge, but there is no dominant coupling into the first waveguide to form an edge state under linear condition [Fig. 1(c2)]. (2) For tilted excitation, however, the beam can easily enter into the edge waveguide under linear condition [Fig. 1(d1)], while it does not efficiently excite the edge waveguide within the 20mm of nonlinear propagation [Fig. 1(d2, d3)]. Simulations to much longer distances beyond the crystal length indicate that the energy of the initial beam will eventually dissipate into the bulk under linear propagation. There is a key difference between trivial and nontrivial lattices under nonlinear propagation for tilted excitation: a distinct edge state persists in the nontrivial lattice, but no edge state exists in the trivial lattice. The underlying mechanism is analyzed below in detail from the nonlinear wave theory.

In the second case, the topological defect is located inside the SSH lattice (Fig. 2). To validate the nontrivial lattice established by laser-writing as shown in Fig. 2(a1), a single probe beam is launched straight into the defect channel which leads to a topological interface state (Fig. 2(a2)). Then, two tilted beams are launched from opposite directions ($k_x = \pm 1.4\pi/a$) to pump the interface defect simultaneously as illustrated in the left panel of Fig. 2. When the two beams are in-phase, light cannot couple into the defect channel in the linear condition (Fig. 2b1), but significantly enhanced coupling into the channel is realized in the nonlinear condition (Fig. 2b2). For comparison, similar experiments were performed in the same lattice under same conditions except for two out-of-phase beams, which cannot couple into the defect channel either under linear or nonlinear excitation conditions [Figs. 2(c1, c2)]. For the linear excitation, topological protection prevents energy flowing into the defect. For the nonlinear excitation, nonlinear interaction of the two out-of-phase beams leads to repulsion from each other. Such a remarkable difference can be seen more clearly from the numerical simulation, where the nonlinearity-induced coupling (Figs. 2(b3)) and "repulsion" (Figs. 2(c3)) is evident. These results show clearly that optical beams from different directions can be pumped into a nontrivial defect channel due to optical nonlinearity under proper excitation conditions.

Now that we have presented our experiment and simulation results which demonstrate nonlinear coupling into topologically protected states, we develop a general theoretical protocol for interpreting dynamics in nonlinear topological systems and employ it for our experiments. Let us assume that the linear component of the index of refraction $V_L(\mathbf{x})$ in Eq. (1) represents a topological photonic lattice, which is characterized by a topological invariant such as the Chern number for 2D lattices or the Zak phase for 1D lattices. The initial excitation is given by $\psi(\mathbf{x}, z = 0)$. Subsequent propagation governed by Eq. (1) gives us the complex amplitude of the electric field $\psi(\mathbf{x}, z)$ along the propagation direction, which in turn modulates the total index potential (linear and nonlinear) for any $z$: $V(\mathbf{x}, z) = V_L(\mathbf{x}) + V_{NL}(\mathbf{x}, z)$. To determine and interpret the topological properties of the dynamically evolving nonlinear system, we use *the total index potential* $V(\mathbf{x}, z)$. The corresponding nonlinear eigenmodes $\varphi_{NL,n}(\mathbf{x}, z)$ and nonlinear eigenvalues $\beta_{NL,n}(z)$ are defined by the equation:

$$(K + V_L + V_{NL})\varphi_{NL,n} = -\beta_{NL,n}\varphi_{NL,n}. \tag{2}$$

We point out that nonlinear eigenmodes and their eigenvalues are a function of the propagation distance $z$, because nonlinear beam dynamics is generally not stationary. In contrast, topological invariants of the linear system are drawn upon the linear eigenmodes $\varphi_{L,n}(\mathbf{x})$ with propagation constants $\beta_{L,n}$, obtained from

$$(K + V_L)\varphi_{L,n} = -\beta_{L,n}\varphi_{L,n}, \tag{3}$$

which are obviously not $z$-dependent. In both cases, $n$ denotes the "quantum" numbers associated with the eigenmode, which can be associated with the Bloch wavevector and the band index for periodic photonic structures.

We emphasize several consequences of this approach: (i) The topological properties depend on the state of the system $\psi(\mathbf{x}, z)$ (this is natural because the system is nonlinear). These properties can be *inherited* from the underlying linear topological system, or they can *emerge* due to nonlinearity (see, e.g., Ref.[24]). The *inherited* and *emergent* topological properties should be distinguished, as elaborated in the Discussion section below. (ii) Topological properties can change along the propagation direction. For example, we envision that for some initial conditions the gap in the nonlinear spectrum $\beta_{NL,n}(z)$ could dynamically close and re-open, leading to topological phase transition driven by nonlinearity. (iii) The evolution of the topological properties depends on the initial condition $\psi(\mathbf{x}, z = 0)$. For a given initial condition, the subsequent dynamics yielding $V(\mathbf{x}, z)$ is unique.

Let us apply the protocol to interpret the dynamics observed in the experiment of Fig. 1 (left panel). The linear SSH lattice with $V_L(x)$ is in the topologically nontrivial regime, which hosts two degenerated edge states as illustrated in Fig. 3(a). The propagation constants of the linear eigenmodes $\beta_{L,n}$ are illustrated in Figs. 3(b-c) (they are plotted in the region $z < 0$ for clarity, although they are $z$-independent); we see two bands corresponding to extended states, while the propagation constants of the localized edge states are in the middle of the gap as expected[41]. They are not at "zero energy" though, simply because we employ the continuous model with experimental parameters. One can achieve the zero energy states by adjusting the bottom of the linear potential through a transformation $V_L(x) \rightarrow V_L(x) + \text{constant}$, but shifting the zero-energy by a constant does not change the physics.

First, we analyze the initial excitation which has the shape of the left edge state (colored red in Fig. 3(a)): $\psi(\mathbf{x}, z = 0) = \sqrt{I_0}\varphi_{L,edge}$; this corresponds to the observation from Figure 1(a2). This linear edge state $\varphi_{L,edge}$ has a typical mode profile of topological characteristic: populating only odd-number waveguides with alternating opposite phase along the SSH lattice[44]. It is convenient to introduce the following quantities: (i) the edge state of the nonlinear system, $\varphi_{NL,edge}(\mathbf{x}, z)$, as the eigenmode of the potential $K + V$, which has the largest overlap with the linear edge state $\varphi_{L,edge}$ defined as $F_{edge}(z) = |\langle\varphi_{NL,edge}|\varphi_{L,edge}\rangle|^2$; (ii) the overlap of the overall complex amplitude $\psi(\mathbf{x}, z)$ with the linear edge state defined as $F_{all}(z) = |\langle\psi(\mathbf{x}, z)|\varphi_{L,edge}\rangle|^2/|\langle\psi|\psi\rangle|^2$. The values of the overlaps $F_{edge}(z)$ and $F_{all}(z)$ are always between 0 and 1 by definition; the former tells us how similar the nonlinear and the linear edge states are, and the latter tells us how much power of the beam is populating the linear topological edge state.

In Figure 3(b), we illustrate the eigenvalue evolution of the nonlinear system $\beta_{NL,n}(z)$ for low nonlinearity (see Supplementary Material for calculation details and parameter values). We see that the bands and the nonlinear eigenvalue of the right edge state (plotted for $z > 0$) are essentially identical to those of the linear spectrum $\beta_{L,n}$ (for comparison $\beta_{L,n}$ is also plotted for $z < 0$, even though it is independent of $z$). However, the nonlinear eigenvalue $\beta_{NL,edge}$ of $\varphi_{NL,edge}$ (for the left edge state) is pushed towards the higher band, although still in the gap[34] The nonlinear spectrum $\beta_{NL,n}(z)$ is almost $z$-independent for this initial excitation. Our calculation shows that in this case $F_{edge}(z) \approx 0.99$, while most of the power populates the left edge state as $F_{all}(z) \approx 0.99$. The inset in Fig. 3(b) shows the profile of the topological linear edge state $\varphi_{L,edge}$, together with that of the nonlinear edge state $\varphi_{NL,edge}$ (at $z = 15$mm). We see that the profile of the nonlinear edge state has the proper oscillations pertaining to the topological edge state, with the amplitude in odd waveguides (starting from the edge waveguide as number one), and opposite phases in neighboring peaks. The edge state has amplitude mainly in the first (edge) waveguide, and then the third waveguide. If the nonlinearity is increased above some threshold value, the nonlinear eigenvalue $\beta_{NL,edge}$ moves across the band to appear above the first band as illustrated in Fig. 3(c). From the mode profiles shown in the inset of Fig. 3(c), we find that the nonlinear edge state is essentially identical to the linear one in the edge waveguide, but it lacks the amplitude in the third waveguide. Such a difference is better seen when we use a larger lattice coupling than the one from experimental parameters.

We conclude that for the initial excitation $\psi(\mathbf{x}, z = 0) = \sqrt{I_0}\varphi_{L,edge}$, when $\beta_{NL,edge}$ is in the gap, the localization is induced by topology, and the nonlinear edge state can be regarded as a topological edge state. When $\beta_{NL,edge}$ is above the upper band (in the semi-infinite gap), the localization is induced by nonlinearity. Even though the mode profile in the edge channel is *inherited* from the linear topological system (see inset in Fig. 3(c)), due to the lack of mode feature in the third waveguide, the nonlinear edge mode should not be characterized as topological when $\beta_{NL,edge}$ is in the semi-infinite gap. A related analysis of similar scenarios can be found in Refs.[30, 59].

Theoretical analysis of the experiments corresponding to tilted excitation in Fig. 1(b2) is more involved, because in this case the dynamics is far from stationary, yet it captures the essence of the theoretical protocol. The beam is launched at $x = 1.2a$, at an angle $k_x = -1.4\pi/a$ towards the edge located at $x = 0$ (see Fig. 3(a)). For this initial excitation, $F_{all}(z = 0) \approx 0$, i.e., at the input of the medium the beam does not excite the linear edge state. Figure 1(b1) is easily understood as $F_{all}(z)$ is $z$-independent in the linear dynamics. Evolution of the nonlinear spectrum $\beta_{NL,n}(z)$ is depicted in Fig. 3(d). First, we note that the band structure (thick blue lines) corresponding to the bulk states is essentially $z$-invariant, and equivalent to that of the linear system. Due to the self-focusing nonlinearity, the dynamics are manifested by the localized modes of $V(x, z) = V_L(x) + V_{NL}(x, z)$; there are quite a few of evolving localized modes of $V(x, z)$ with eigenvalues $\beta_{NL,n}(z)$ indicated by dotted blue lines in Fig. 3(d). We focus only on the nonlinear edge state $\varphi_{NL,edge}$, and its eigenvalue $\beta_{NL,edge}$ plotted in Fig. 3(d) with a solid red line. From Figs. 3(e) and (f), which illustrate $F_{all}(z)$ and $F_{edge}(z)$, respectively, we see that dynamics can be divided in three stages. More specifically, a sudden drop of $F_{edge}(z)$ at $z = 5$mm indicates the end of the first stage, while a sudden increase at $z = 11$mm indicates the end of the second stage of the dynamics (see Fig. 3(f)). In the first stage (shaded magenta in Figs. 3(d) and (f)), the launched beam travels towards the edge, and the edge state is not populated as $F_{all}(z) \approx 0$; consequently $\beta_{NL,edge}$ is in the gap (see the left red line in Fig. 3(d)), and $F_{edge}(z)$ is close to unity. In the second stage (shaded gray) when the beam is at the edge, the linear edge state gets populated and $F_{all}(z)$ increases. In this stage the beam strongly perturbs the local structure of the lattice at the edge, as can be seen from the drop of $F_{edge}(z)$ in Fig. 3(f), which means that none of the nonlinear localized states are similar to $\varphi_{L,edge}$ (thus none of the nonlinear eigenvalues is colored red in Fig. 3(d) in the second stage). In the third stage (shaded green), a large portion of the beam gets reflected, but about 30% of the beam gets trapped into a localized edge state: $F_{all}(z) \approx 0.3$, as shown in Fig. 3(e). There is a well-defined nonlinear edge state, with eigenvalue $\beta_{NL,edge}$ above the first band, not in the gap, as indicated by the right red line in Fig. 3(d). The profile of this nonlinear edge state is mostly inherited from the topology of the linear structure as can be seen from the inset of Fig. 3(e) and the overlapping $F_{edge}(z) \approx 0.98$ shown in Fig. 3(f); however, it lacks the topological mode feature in the third waveguide. We conclude that the localization is dominantly nonlinear. We should emphasize that the linear edge state does not continuously transform into the nonlinear edge state during propagation, because of the strong deformation of the lattice in the second stage of the dynamics. After this distortion, one of the localized states from stage two re-emerges as the new nonlinear localized edge state in stage three, as can be traced by following the nonlinear eigenmodes alongside with the $F$-functions plotted in Figs. 3(d)-(e).

The details of the theoretical analysis corresponding to Fig. 1 right panel (for the SSH lattice in the topologically trivial regime), and Fig. 2 (for excitation of the topological defect with two beams) are shown in the Supplementary Material and summarized here. Results in Fig. 1 right panels can be interpreted as follows. All the linear modes are extended, as the SSH lattice is in the topologically trivial regime. The beam initially excites many of these states. In the linear regime illustrated in Figure 1(d1), the beam approaches the waveguide at the edge, and then travels along the edge for the length of the crystal. In other words, for short propagation distances (smaller than the length of the crystal), the phases of all linearly excited (extended) modes add such that the intensity of the beam populates the waveguides close to the edge in Figure 1(d1). However, for a very long propagation distance, due to dephasing of the excited bulk modes, the beam will spread into the lattice. In the nonlinear case corresponding to Figs. 1(d2) and (d3), the nonlinearity creates evolving localized states, which are not related to the topological origin, as none of the nonlinear modes resemble the linear topological edge state. In fact, in this trivial lattice structure, the localized modes arise purely due to the nonlinear index change, as in the case typically with optical solitons. A light beam forms a few self-trapped filaments around these states and evolves in this fashion for the propagation distance of the crystal length in experiment. As the initial excitation is not at the edge, the location of the self-trapped filaments is also not at the edge as illustrated in Figs. 1(d2) and (d3).

Regarding the excitation of the defect mode with two beams at opposite angles, when the beams are in-phase, there are again three stages of the dynamics, which are equivalent to those presented in Figs. 3(d)-(f). In the first stage, the beams travel towards the linear defect channel in the center of the lattice; the defect state is not yet populated, and its eigenvalue is in the gap. Many evolving nonlinear localized states arise due to nonlinearity but not topology. In the second stage, the linear defect state starts to get populated, but the lattice is distorted locally due to nonlinear action, so none of the nonlinear states are similar to the linear defect state. In the third stage, some of the incident light (about 20-30% for the parameters used here) is trapped in the defect state, while the rest is repelled. There is a well-defined nonlinear defect state with its profile in the defect channel inherited from the linear defect state, with a nonlinear eigenvalue emerged above the first band. Thus, conceptually an identical scenario to that described in Figs. 3(d)-(f) occurs. The difference is that the defect state can now be coupled from both sides, which could be extended to coupling light from all directions in a 2D SSH-type system, leading to a nonlinear "tapered" topological waveguide. Such potential applications certainly merit further research.

When the two incident input beams at opposite angles are out-of-phase, again there are three stages of the dynamics analogous to the ones presented before (see Supplementary Material). However, the linear defect state does not get populated by any of them. The eigenvalue of the nonlinear defect state is within the gap in the first and the third stage of the dynamics. In the second stage when the light is close to the defect state, the lattice structure is distorted and none of the nonlinear localized states is very similar to the linear defect state. In fact, the two beams stay away from each other and the defect in this case is related to nonlinear interaction of out-of-phase soliton-like beams rather than to topology.

## Discussion

The interplay of nonlinearity and topology is somewhat analogous to the interplay of locality and globality, as most of the studied optical nonlinearities are local, and the topology describes global properties of a system. In order to analyze nonlinear topological systems, one must find an appropriate way to connect the local and global properties of the underlying systems. The introduced theoretical protocol does just that: it takes the total change in the index potential $V(\mathbf{x}, z) = V_L(\mathbf{x}) + V_{NL}(\mathbf{x}, z)$ (which includes the nonlinear term), and analyzes the topological properties of a nonlinear system.

Our theory is designed to unravel non-stationary dynamics, which is at the essence of nonlinear coupling presented here. In this case, the potential $V(x,z) = V_L(x) + V_{NL}(x,z)$ evolves along $z$ ($z$ is the "time" in our system), and topological quantities can in principle change during evolution. In the specific lattice system studied above, the gap in the nonlinear spectrum $\beta_{NL,n}(z)$ does not close at any $z$, and the bands remain pretty much intact in the presence of the nonlinearity. However, we have observed that the interplay of nonlinearity and topology can couple light into the topological edge state of the linear system, which is inadmissible for entirely linear dynamics (e.g., see Figure 1(b2) and Figures 3(d)-(f)). When this happens, we can identify the nonlinear edge mode $\varphi_{NL,edge}$, which has inherited the profile of the linear edge mode $\varphi_{L,edge}$ in the edge channel, quantified by $F_{edge}(z) \approx 0.98$ after the nonlinear coupling has occurred, but it lacks the amplitude in the third waveguide. Thus, for a high nonlinearity, the eigenvalue of $\varphi_{NL,edge}$ moves outside the gap (see Figure 3(c)), the edge mode is dominated by nonlinearity, but with some features inherited from the linear topological edge mode; for a low nonlinearity, its eigenvalue stays inside the gap, so it is dominantly a topological mode. For the other initial conditions studied in the experiment, presented in Figure 2, the interplay of topology and nonlinearity is conceptually the same.

Let us comment on the calculation of topological invariants in finite nonlinear lattices. Topological invariants for periodic lattices, the Chern number for 2D lattices and the Zak phase for 1D lattices, are calculated for an infinite periodic system by integrating over the Brillouin zone[2]. These invariants in a finite lattice are manifested by edge states (edge modes), such as the ones in the SSH model. When we deal with a finite nonlinear system, we cannot straightforwardly use the formulae for calculating the Chern number and the Zak phase for the infinite periodic systems. This problem was already addressed in the literature, e.g., see Refs.[63, 64], where the so-called Bott index was calculated. Here we developed an approach which explores the nonlinear eigenmodes and compares them with the pertinent linear eigenmodes. This relies on the following facts: (i) the Zak phase (and the Chern number) is calculated from the eigenmodes, and (ii) for the linear system, the Zak phase is well known in the topologically trivial or in the nontrivial regime. Thus, we have focused on how the eigenmodes change when we introduce the nonlinearity. This is quantified in the overlaps of the linear and nonlinear modes and visualized in the mode profiles and the positions of their eigenvalues in the spectrum.

Before closing, let us discuss the distinction between the *inherited* and *emergent* nonlinear phenomena. If the underlying nonlinear system is topological, such distinction is manifested in topological invariants, pertinent to edge modes of the system, and perhaps in other quantities. During nonlinear evolution, some of the local quantities such as the edge modes can be modified by the nonlinearity, without closing the gap or changing the topological invariants. If these modified nonlinear modes are similar by some measure (such as the quantity $F_{edge}$ used here) to the modes of the underlying linear system, we say that their properties are *inherited*. However, if the underlying linear system is initially topologically trivial, under some conditions it may happen in such a way that the nonlinear dynamics changes the topological invariant and turns the system into topologically nontrivial regime. Because the action of the nonlinearity is normally local, for such a scenario to occur, it appears that the excitation should be extended. For this type of scenario, which is in principle possible, we say that the topological properties of the nonlinear system are *emergent*, because they are not present in the corresponding linear system. Although in the particular setting employed in this study, emergent nonlinear phenomena such as band inversion and topological phase transition have not been observed, we envision that they should exist in some nonlinear topological systems.

## Materials and Methods

Our experiment method for laser-writing the 1D SSH photonic lattice is shown in Fig. 4, where Fig. 4(a) illustrates the idea to establish the superlattice by overlapping two periodic index potentials of different periods[43], and Fig. 4(b) depicts the experimental setup. The two periodic potentials are denoted by the dashed curves in Fig. 4(a), which are written into the nonlinear SBN crystal one after another due to optically induced local index change. Superposition of these two potentials leads to the SSH lattice (solid curve), where the coupling of neighbor sites can be fine-tuned by shifting their relative position. When the lattice is terminated at the strong-coupling "bond" denoted by $t$, it corresponds to the nontrivial case shown in the left panels of Fig. 1, since in this case the intra-cell coupling is weaker than the inter-cell coupling. The opposite case is when the lattice is terminated at the weak-coupling "bond" denoted by $t'$, which represents a trivial lattice shown in the right panels of Fig. 1. Since the cw-laser writing technique[60] is used to induce the potential one by one, the lattice edge (and interface) can be readily reconfigured by this method.

In the setup of Fig. 4(b), the upper and lower paths correspond to the lattice-writing and probing beams, respectively. A collimated laser beam (with wavelength 532 nm and power 100mW) illuminates a programmable spatial light modulator (SLM), which alternatively generates the writing and probing beams. In the writing path, the beam exiting the SLM is collimated and spatially filtered with a narrow single-slit, and then further compressed into a narrow stripe beam with an FWHM about 10μm by a pair of cylindrical lenses, so it is long enough to cover the entire 20 mm-long SBN:61 crystal. Its input position to the crystal is precisely controlled by the SLM. Through a multi-step writing process in the biased crystal (with applied field 240kV/m), the desired SSH lattice is established with a lattice constant of 38μm. Thanks to the "memory" effect of the photorefractive crystal, such an index lattice remains intact for more than one hour, enough time for measurement of the beam dynamics. In the lower path, the probe (stripe) beam is launched into the lattice, and its input size, position and direction can all be adjusted by the SLM. In addition, the probe beam can undergo linear or nonlinear propagation through the lattice, depending on whether a proper bias field is applied or not[62]. The CCD camera in the writing beam path is used to examine the position of the stripe writing beam, and the other CCD is used to monitor the input and output of the probe beam propagating through the lattice. In order to image a particular SSH photonic lattice after it is written, a single stripe beam is launched into the crystal to probe the waveguides one by one, and then all guided outputs of the probe beam are superimposed to display the lattice structure of Figs. 1(a1) and (c1).

## Conclusion

In conclusion, we have established trivial and nontrivial photonic SSH lattices by direct cw-laser writing in a bulk nonlinear crystal, and thereby experimentally demonstrated nonlinearity-induced coupling of light into a topological edge state. In particular, we have shown that two optical beams from different directions can couple into (stay away from) a nontrivial defect channel under nonlinear (linear) excitation upon collision. We have developed a theoretical protocol to explain the observed dynamics in this lattice system. Our theory shows that, by nonlinear excitation of bulk modes, depending on the input power (i.e., the strength of the nonlinearity), the trapped light beam can evolve into a nonlinear edge mode with profile featuring the topological edge state fully inherited from underlying linear system. These features exemplify the interplay of topology and nonlinearity in topological nontrivial systems. The protocol presented in this work is general, applicable not only for non-stationary and dynamically evolving systems such as the one studied here, but also for other systems besides the SSH lattices and even beyond the photonic platform.

For future research we envision many fundamental issues could arise from systems with *emergent*, rather than solely *inherited*, nonlinear topological phenomena, where nonlinear dynamics can close and open the gap and induce topological phase transition. The toolkit for such studies has been presented here. Our results may bring about insights and advances in nonlinear control of topological quantum states in similar systems[28, 36, 54, 55], as well as to photonic parity-time-symmetric systems where the excitation can be tuned by nonlinearity[48].

## Acknowledgements

We thank U. Peschel, K. Makris and D. Li for discussion and assistance. This research is supported by the National Key R&D Program of China under Grant (No. 2017YFA0303800), the National Natural Science Foundation (11922408, 91750204, 11674180), PCSIRT, and the 111 Project (No. B07013) in China. D.J. and H.B. acknowledge support in part by the Croatian Science Foundation Grant No. IP-2016-06-5885 SynthMagIA, and the QuantiXLie Center of Excellence, a project co-financed by the Croatian Government and European Union through the European Regional Development Fund - the Competitiveness and Cohesion Operational Programme (Grant KK.01.1.1.01.0004). D.S. has been supported by the Australian Research Council (DE19010043). D.L. is supported by the Institute for Basic Science in Korea (IBS-R024-Y1). L.S. acknowledges support from the Russian Foundation for Basic Research (grant No 19-52-12053).

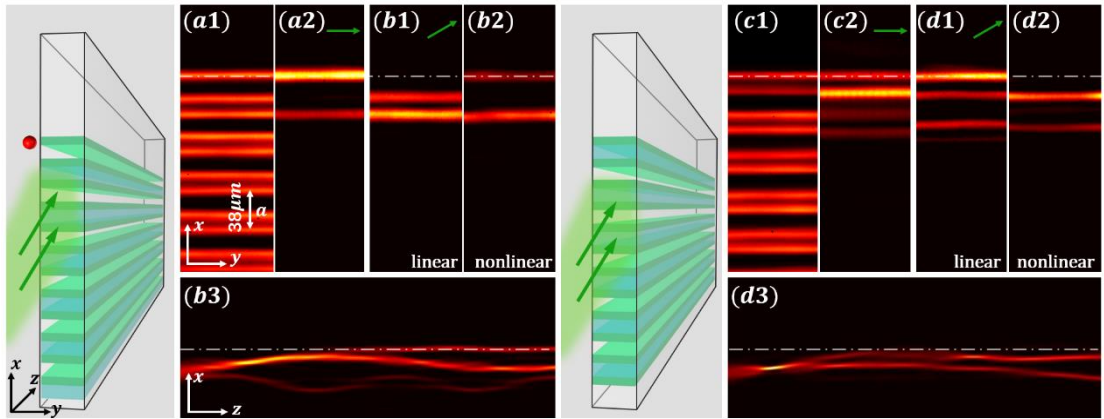

**Fig. 1 Comparison of edge excitation between topological trivial and nontrivial SSH lattices.** The two illustration figures show tilted excitation (green arrows) for nontrivial (left) and trivial (right) lattices, where the red dot marks the position of the nontrivial edge. (a1-b2) Experiment results obtained with the nontrivial lattice, where (a1) shows the written SSH waveguide lattice examined by a probe beam, (a2) shows the output of a topological edge state under normal (straight) excitation, (b1) and (b2) show the outputs with a tilted beam ($k_x = -1.4\pi/a$) under linear and nonlinear cases. (b3) Simulation results show side-view (up to crystal length of 20 mm) of the beam dynamics under nonlinear excitation. The right panels (c1-d3) have the same layout as the left ones except that the results are obtained with the trivial lattice. White dash-dotted line marks the edge position of the SSH lattice in all figures.

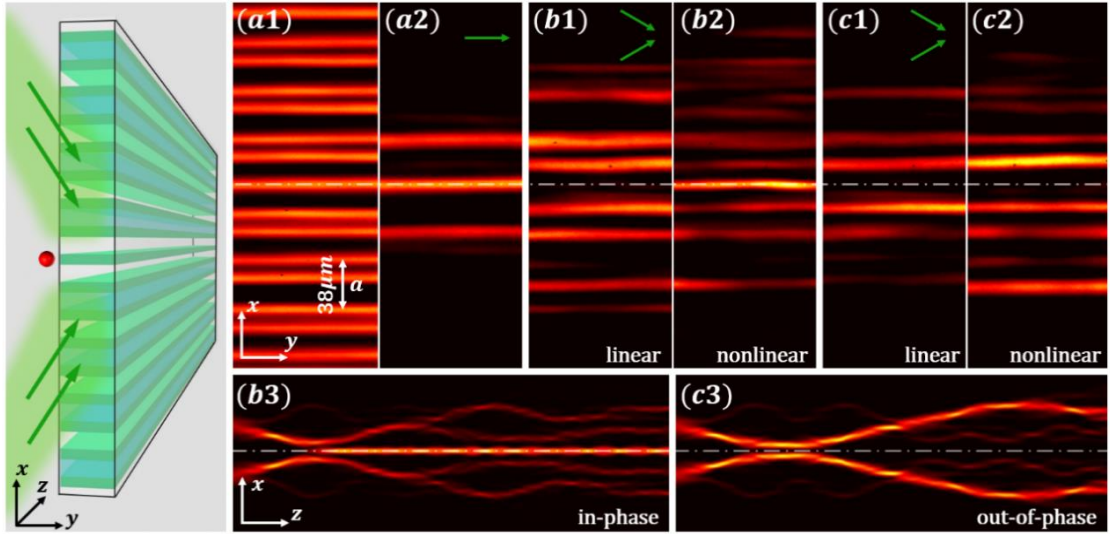

**Fig. 2 Nonlinearity-induced coupling and "escaping" of topological interface state.** The illustration figure in the left shows two-beam tilted excitations (green arrows) of the topological defect from opposite directions. Top panels are from experiment, where (a1) shows the cross section of the lattice, (a2) shows the output of a topological interface state under single-beam (straight) excitation, (b1) and (b2) show outputs of two tilted in-phase beams ($k_x = \pm 1.4\pi/a$) under linear and nonlinear excitation conditions. Bottom panel (b3) is from simulation, showing side-view of beam dynamics (up to the length of 40 mm) under nonlinear excitation. The right panels (c1-c3) have the same layout as (b1-b3) except that the defect is excited with two tilted out-of-phase beams. White dash-dotted line marks the position of nontrivial interface defect channel in the SSH lattice in all figures.

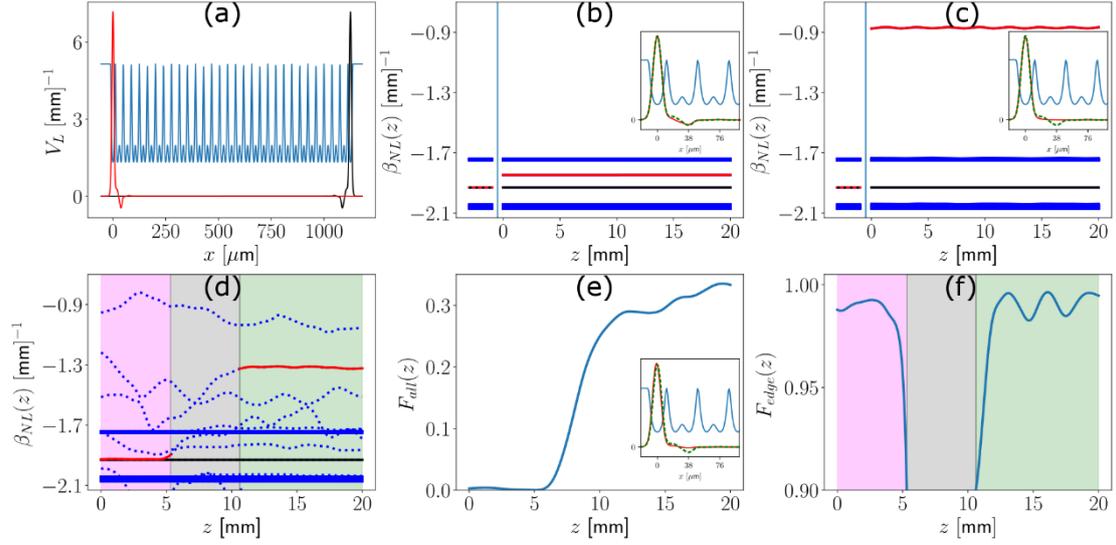

**Fig. 3 Nonlinear evolution of eigenvalues and coupling to the edge states in topological nontrivial SSH lattices.** (a) Two linear edge states (red and black) found in the SSH lattice (dark blue) used in our theoretical analysis. (b, c) Band structure and nonlinearity-induced eigenvalue shifting under normal (straight) excitation condition at a low (b) and high (c) nonlinearity; the insets show the linear topological edge mode (green dashed line) and the nonlinear edge mode (red solid line). The evolving nonlinear eigenvalues $\beta_{NL,n}(z)$ are shown for $z > 0$. For comparison, the linear spectrum $\beta_{L,n}$ is shown for $z < 0$. Red line is the eigenvalue of the (left) nonlinear edge mode, and black line corresponds to the (right) linear edge mode, which is not excited. Thick blue lines are the bands. (d) Nonlinear eigenvalue evolution under tilted excitation condition (corresponding to left panels of Fig. 1). Red (black) line depicts the nonlinear (linear) edge eigenvalue as in (b, c), while individual blue dotted lines correspond to nonlinear localized states not inherited from the linear topological edge states. Three stages of dynamics described in the text are denoted with magenta, grey, and green shaded regions, respectively. (e) The overlap of the whole beam with the linear edge state $F_{all}(z)$. (f) The overlap of the linear and the nonlinear edge mode $F_{edge}(z)$ at three stages of the evolution. See text for details.

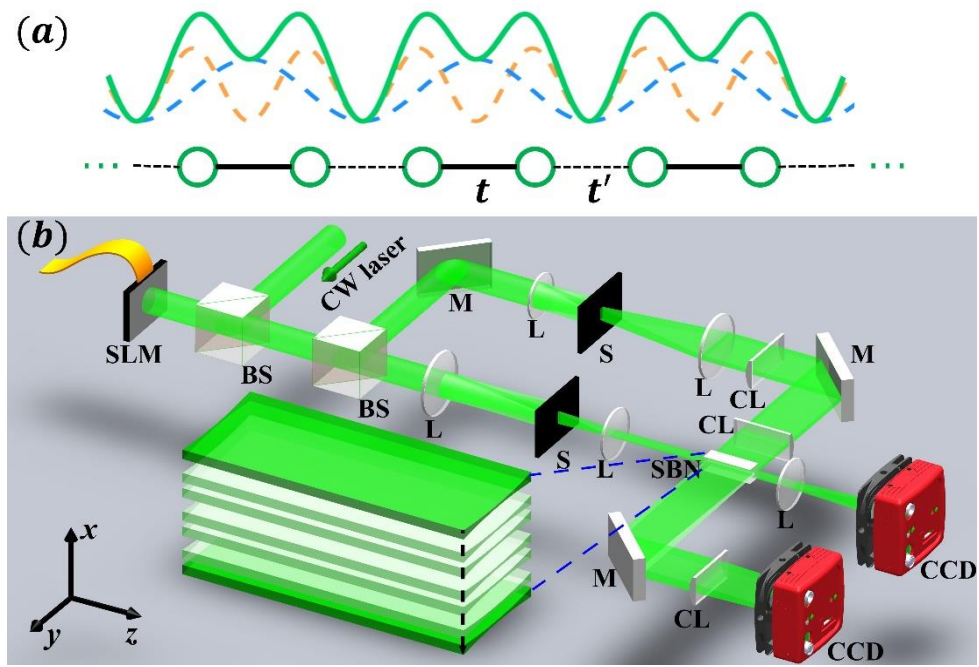

**Fig. 4 Experimental scheme for laser-writing photonic SSH lattices in a nonlinear crystal.** (a) Illustration of the SSH model, where $t$ and $t'$ represent strong and weak coupling, respectively. Green solid curve shows the SSH lattice superimposed from two periodic lattices depicted by dashed curves. (b) Experimental setup for writing and probing the SSH lattice. SLM: spatial light modulator; BS: beam splitter; M: mirror; L: circular lens; S: single slit; CL: cylindrical lens; SBN: strontium barium niobite crystal; CCD: camera. Upper path is for the lattice-writing beam (ordinarily-polarized), and lower path is for the probe beam (extraordinarily-polarized). The lattice structure in the crystal is zoomed-in and illustrated in the lower inset.

# Supplementary Material:

**Details of theoretical analysis**

**1. Edge excitation with a single beam in nontrivial SSH lattices**

Details of the theoretical analysis for Fig. 3 in the main text are presented here, corresponding to edge excitation with a single input beam. The total potential (linear and the nonlinear terms), corresponding to the photorefractive medium used in experiments is

$$V(x,z) = \frac{k_0}{n_0} \frac{\Delta n}{1+i_L(x)+i_{NL}(x,z)}, \quad (1)$$

where $n_0$=2.35, $k_0 = \frac{2\pi n_0}{\lambda}$, $\lambda = 532$ nm, $\Delta n = 4.36 \times 10^{-4}$;

$$i_L(x) = N_L \left[ 0.8 \left( \cos(\frac{\pi(x-c)}{a}) \right)^2 + \left( \cos(\frac{\pi(x-c-a/4)}{a/2}) \right)^2 \right], \quad (2)$$

where $a = 38$ μm is the lattice constant, $c$ is offset constant so that the left edge is centered at $x = 0$, and the normalization $N_L$ is such that the lattice amplitude (the maximum value of $i_L(x)$) is 2.88. The nonlinear contribution is

$$i_{NL}(x) = \gamma |\psi(x,z)|^2, \quad (3)$$

where $\gamma$ tunes the intensity of the beam without changing its shape or the bias field (and therefore the strength of the nonlinear index change). The quantities $i_L(x)$ and $i_{NL}(x)$ are dimensionless. For the SBN crystal used in the experiments, the nonlinear index change parameter is $\Delta n = -n_0^3 r_{33} E_0/2$, where $r_{33}$ is the effective electro-optic coefficient of the SBN crystal, and $E_0$ is the bias field. The strength and the sign of the bias field determines the strength and the sign of the nonlinearity. When the bias field is zero, i.e., $E_0 = 0$, the system is linear.

The initial states for Fig. 3(b) and (c) are identical in shape to the linear edge state, $\psi(\mathbf{x}, z = 0) = \sqrt{I_0} \varphi_{L,edge}$. For low nonlinearity [Fig. 3(b)], we choose $\gamma \max|\psi(\mathbf{x},0)|^2 = 0.30$, and for strong nonlinearity [Figs. 3(c)-(f)], we set $\gamma \max|\psi(\mathbf{x},0)|^2 = 12.5$. At the critical value of $\gamma \max|\psi(\mathbf{x},0)|^2 = 0.72$, the nonlinear eigenvalue $\beta_{NL,edge}$ moves from inside the gap [Fig. 3(b)] to above the first band [Fig. 3(c)].

The shape of the initial state for simulations (mimicking the input broad beam for lattice excitation used in experiment) in Figs. 3(d)-(e) is given by

$$\psi(x, z = 0) = \sqrt{I_0} \exp\left(\frac{-i1.4\pi x}{a}\right) \exp\left(-\frac{(x - 45.6 \text{ μm})^2}{(22.8 \text{ μm})^2}\right).$$

(4)

The strength of the nonlinearity is set to $\gamma I_0 = 4.10$.

## 2. Interface excitation with two beams in nontrivial SSH lattices

Here we provide theoretical analysis of experiments and numerical simulations presented in Fig. 2 of the main text. The theoretical protocol has already been applied in the main text to obtain Fig. 3 and explain the experimental results of Fig. 1.

The outline of Fig. S1(a)-(c) is identical to that of Fig. 3(a)-(c) from the main text, except now it is for the SSH lattice with an interface defect located in the center of the lattice. In Fig. S1(a) we show the linear SSH lattice which has three localized states in the band gap: the left (red) and the right (black) edge states, and the interface (magenta) defect state. When the input excitation is chosen to match the shape of the (linear) defect state, $\psi(\mathbf{x}, z = 0) = \sqrt{I_0}\varphi_{L,defect}$, for low nonlinearity $\gamma \max|\psi(\mathbf{x}, 0)|^2 = 0.312$, the eigenvalue of the defect state $\beta_{NL,defect}(z)$ is shifted towards the first band due to nonlinearity but remains inside the gap, as shown in Fig. S1(b). We see that $\beta_{NL,defect}(z)$ essentially stays constant during the propagation. When the nonlinearity is increased to a threshold value, $\beta_{NL,defect}(z)$ is found to be above the first band, as illustrated in Fig. S1(c) for $\gamma \max|\psi(\mathbf{x}, 0)|^2 = 13.0$.

Now we discuss the scenario in which two tilted beams are sent from opposite angles towards the interface defect site, as described in Fig. 2 of the main text. In this example we use the same parameters as that for Figs. 3(d)-(f) of the main text, that is, the same structure of the input beams described by Eq. (4) above (one from left, and the other from right of the defect). In Figs. S1(d)-(f) we show results for the case when two beams are in phase, and corresponding results for the out-of-phase beams are shown in Figs. S1(g)-(i). The outline of Figs. S1(d)-(f) [and S1(g)-(i)] is identical to that of Fig. 3 (d)-(f) from the main text, with three shaded regions (magenta, gray, and green) corresponding to three different stages of nonlinear beam dynamics.

When two beams are in phase, we again have three stages of the dynamics analogous to those from Fig. 3 in the main text. The evolution of the nonlinear eigenvalue spectrum $\beta_{NL,n}(z)$ is presented in Fig. S1(d). In the first stage (shaded in magenta), the wavepacket has not yet arrived at the defect site, and the eigenvalue $\beta_{NL,defect}(z)$ of the nonlinear defect state $\varphi_{NL,defect}$ is the same as in the linear case, as denoted by the solid magenta line. In addition to two edge states and the defect state elaborated so far, we observe additional localized states which appear due to nonlinearity (and not topology of the linear lattice); their nonlinear eigenvalues are indicated with blue dotted lines. In the first stage, the beam does not excite the defect state, i.e., the overlap $F_{all}(z) = |\langle\psi(x,z)|\varphi_{L,defect}\rangle|^2/|\langle\psi|\psi\rangle|^2$ is close to zero, as illustrated in Fig. S1(e). Consequently, in this first stage $F_{defect}(z) = |\langle\varphi_{NL,defect}|\varphi_{L,defect}\rangle|^2$ is approximately one, as shown in Fig. S1(f). In the second stage (shaded grey), the two beams have arrived at the defect site, which changes significantly the local structure of the lattice. None of the nonlinear eigenmodes are similar to the linear defect state $\varphi_{L,defect}$, as can be seen from the drop of $F_{defect}(z)$ illustrated in Fig. S1(f). In this stage, the linear defect state becomes populated, see Fig. S1(e). In the third stage (shaded green), the wavepacket is partially reflected, but a part of it stays at the defect site (about 20-30% as can be seen from Fig. S1(e)). The nonlinear defect state $\varphi_{NL,defect}$ has again a great overlap with the linear defect state $\varphi_{L,defect}$ [Fig. S1(f)], but its propagation constant $\beta_{NL,defect}(z)$ is now above the first band [Fig. S1(d)].

Figures S1(g)-(i) illustrate results obtained for the two beams initially out of phase. Evolution of the nonlinear spectrum is shown in Fig. S1(g). In sharp contrast to the in-phase case, the defect eigenvalue $\beta_{NL,defect}(z)$ remains inside the gap at all stages, as seen from Fig. S1(g). The overlap of the whole beam with the linear defect state $F_{all}(z) \approx 0$ at all times, as seen from Fig. S1(h), indicating that coupling to the defect state does not occur. From Fig. S1(i), we see that as the beams approach the defect, $F_{defect}(z)$ decreases. This occurs due to the change of the local index of refraction, i.e., the local distortion of the lattice. In this case the coupling does not occur, and $F_{defect}(z)$ revives after the beams are repelled from each other at the defect.

### 3. Edge excitation with a single beam in trivial SSH lattices

For completeness and direct comparison, in Fig. S2 we present a detailed theoretical analysis corresponding to the excitation of the trivial SSH lattice; pertinent experiments are presented in the right panel of Fig. 1 in the main text. The outline of Fig. S2, and the parameters used such as the nonlinearity strength, are identical to those for Fig. 3 in the main text.

In Fig. S2(a) we show the trivial SSH lattice used in the simulations; there are no topological edge or defect states. In Fig. S2(b,c) we show dynamics of the nonlinear eigenvalues for the initial excitation which has the shape of the topological edge state of the nontrivial SSH lattice: $\psi(\mathbf{x}, z = 0) = \sqrt{I_0}\varphi_{L,edge}$ (identical initial condition as for Fig. 3(b,c)). We observe that dynamics of the bands is practically $z$-independent (thick blue lines); the bands are the same as for the underlying linear system. In Fig. S2(b) we see that there are two nonlinear localized states, one with the eigenvalue in the semi-infinite gap above the first band (solid red line), and the other with the eigenvalue in the gap (dotted red line). In Fig. S2(c) we again see two localized states, with eigenvalues above the first band. For this initial excitation, most of the beam power is present in the two eigenmodes induced by the nonlinearity (no feature of topology is present), which evolve along the propagation axis.

In Fig. S2(d-f) we show results for the excitation at an angle towards the edge of the trivial lattice. Figure S2(d) illustrates dynamics of nonlinear eigenvalue spectrum. We see the bands which are $z$-independent (thick blue lines). Dynamics is again manifested in the evolution of nonlinearly excited localized eigenmodes, whose propagation constants are illustrated with dotted blue lines. During the evolution, none of them resemble the topological linear defect state of the nontrivial SSH lattice, which follows from Fig. S2(f) displaying $F_{edge}(z)$; we only see an occasional rise of the overlap above 70%; this occurs due to the nonlinear mode beating, but it does not persist and cannot reach high values to be related to the topological edge states in any way (as compared to 98-99% overlap observed in Fig. 3 of the topologically nontrivial lattice reported in the main text); it does not make sense to display smaller overlaps as they carry no meaning subject to interpretation. From Fig. S2(e) we see that the overlap of the beam with the topological edge state of the linear system is small at all times, indicating that this is not a topological phenomenon as those occurring in the nontrivial lattice, but rather it is due to the nonlinearity.

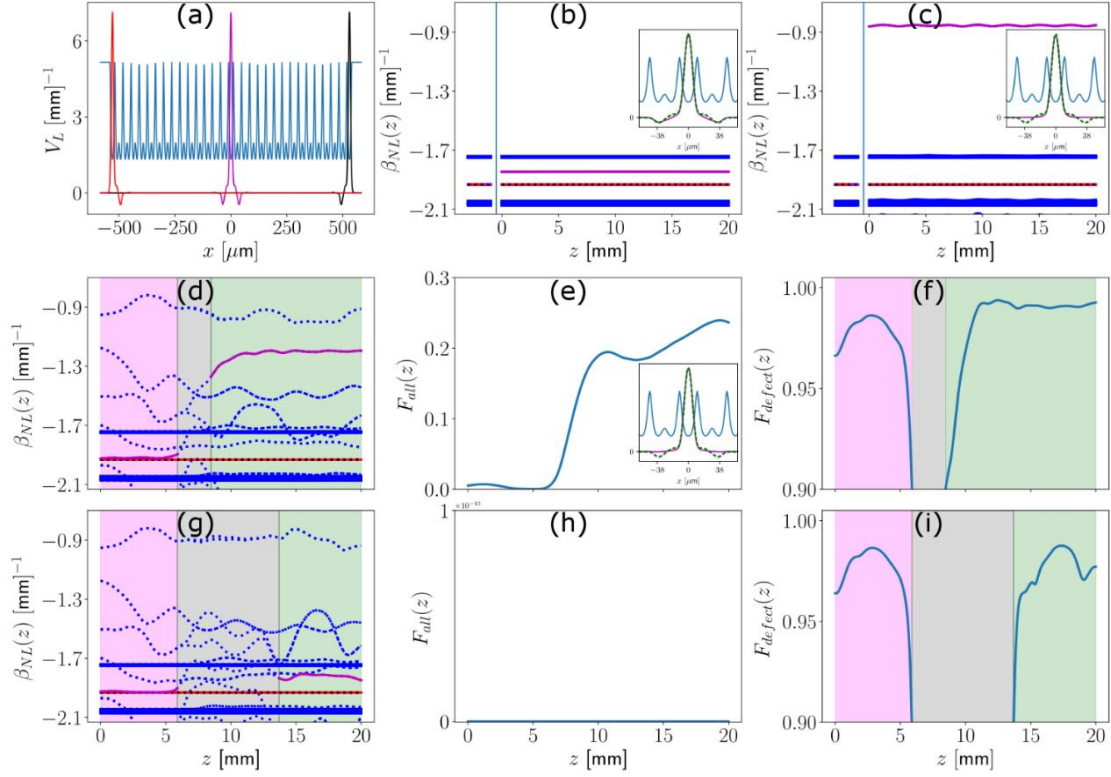

**Fig. S1 Dynamics of nonlinear eigenvalues and the overlap between the linear and nonlinear defect states.** (a) The linear SSH lattice (dark blue lines) with one defect state in the center (solid magenta line) and two edge states (solid red and black lines). (b, c) Nonlinear eigenvalues $\beta_{NL,n}(z)$ for $\psi(\mathbf{x}, z=0) = \sqrt{I_0}\varphi_{L,defect}$ at low (b) and high (c) nonlinearity; for comparison, the linear spectrum $\beta_{L,n}$ is shown for $z < 0$. Magenta line depicts the nonlinear eigenvalue $\beta_{NL,defect}(z)$ of the interface defect mode $\varphi_{NL,defect}$. Red and black lines correspond to the edge modes which are not populated. Thick blue lines are the bands. The insets show the linear topological mode (green dashed line) and nonlinear defect mode (magenta solid line) (d)-(f) Results for two beams launched in phase towards the defect. (d) Dynamics of the nonlinear eigenvalues $\beta_{NL,n}(z)$; the color notations for the defect mode and the two edge modes are the same as in (b) and (c). Dotted blue lines indicate the nonlinear eigenvalues of modes localized solely due to the nonlinearity. (e) The overlap of the whole beam and the linear defect mode $F_{all}(z)$. (f) The overlap of the nonlinear defect mode and the linear defect mode $F_{defect}(z)$. (g-i) Dynamics for the two beams initially out of phase. The outline is identical to that for (d-f). Three shaded regions (magenta, gray, and green) in (d, f, g, i) correspond to three different stages of nonlinear beam dynamics. See text for details.

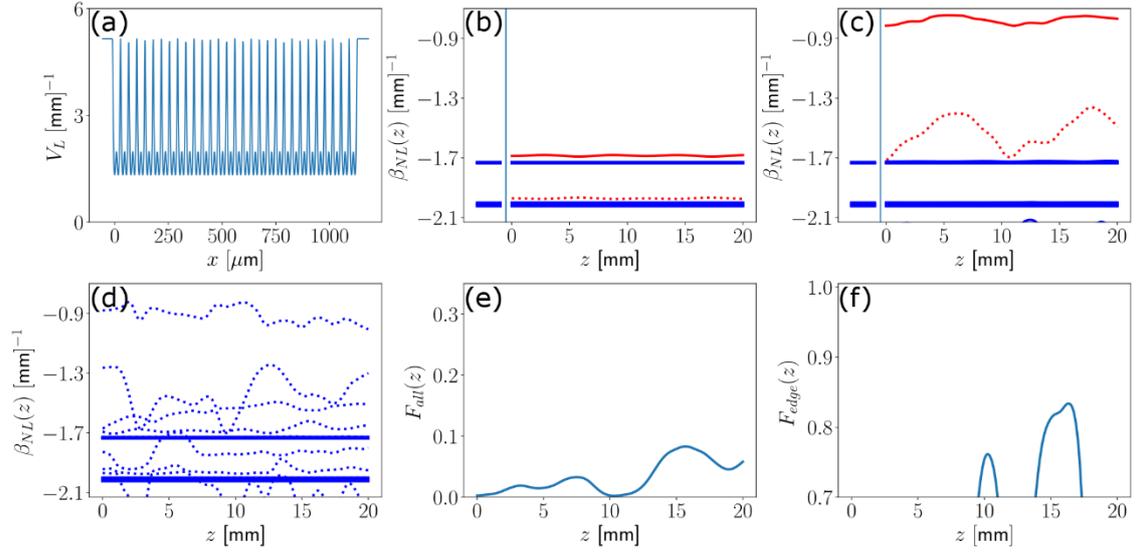

**Fig. S2 Dynamics of nonlinear eigenvalues for excitations of the SSH lattice in the topologically trivial regime.** (a) The linear SSH lattice (dark blue lines) in the topologically trivial regime. (b, c) Nonlinear eigenvalues $\beta_{NL,n}(z)$ for $\psi(\mathbf{x}, z = 0) = \sqrt{I_0}\varphi_{L,defect}$ at low (b) and high (c) nonlinearity. Red solid and red dotted lines correspond to the nonlinear localized modes. These modes are purely nonlinear and not related to the topology of the SSH lattice. (d-f) Dynamics for a beam launched at an angle towards the edge. (d) Evolution of nonlinear eigenvalues $\beta_{NL,n}(z)$. Thick blue lines correspond to the bands, and dotted blue lines correspond to the (purely) nonlinear localized states. (e) The overlap of the whole beam with the linear edge mode of the topologically nontrivial SSH lattice $F_{all}(z)$; small values indicate that the nonlinearly excited modes are not related to linear topological states. (f) None of the nonlinear localized modes resemble the structure of the linear edge mode of the topologically nontrivial SSH lattice, as seen from $F_{edge}(z)$; occasional and accidental overlaps above 70% arise from the nonlinear dynamics of mode beating, but they do not persist and cannot reach high values of 98-99% overlaps presented in Fig. 3 of the main text.